\newcommand{\teff}{$T_{\rm eff}$}
\newcommand{\logg}{$\log\,g$}
\newcommand{\vsini}{$v\,\sin i$}
\newcommand{\change}{}

\documentclass{aa}
\usepackage{graphicx}

\begin{document}
\headnote{Letter to the Editor}

\title{Magnetic fields in Herbig Ae stars
\thanks{Based on observations obtained at the European Southern Observatory, 
Paranal, Chile (ESO programme No.~072.D-0377)}}

\author{S. Hubrig\inst{1}\and M. Sch\"oller\inst{1}
\and R.\,V. Yudin\inst{2,3}}

\institute{European Southern Observatory, Casilla 19001, Santiago 19, Chile
\and Central Astronomical Observatory of the Russian Academy of Sciences at 
Pulkovo, 196140 Saint-Petersburg, Russia
\and 
Isaac Newton Institute of Chile, St.-Petersburg Branch, Russia
}

\date{Received xx/Accepted yy}

\offprints{S. Hubrig}

\authorrunning{Hubrig et~al.}

\abstract{
Herbig Ae stars are young A-type stars in the pre-main
sequence evolutionary phase with masses of $\sim$1.5--3\,M$_\odot$.
They show rather intense surface activity (Dunkin et~al.\ \cite{Du97})
and infrared excess related to the presence of circumstellar disks.
Because of their youth, primordial magnetic fields inherited from the parent molecular
cloud may be expected, but no direct evidence 
for the presence of magnetic fields on their surface, except in one
case (Donati et~al.\ \cite{Do97}), has been found until now.
Here we report observations of optical circular polarization with FORS\,1 at the
VLT in the three Herbig Ae stars HD\,139614, HD\,144432 and HD\,144668.
A definite longitudinal magnetic field at 4.8\,$\sigma$ level, 
$\left<{\cal B}_z\right>$=$-$$450\pm93$\,G, has been detected
in the Herbig Ae star HD\,139614.
This is the largest
magnetic field ever diagnosed for a Herbig Ae star.
A hint of a weak magnetic field is found in the other two Herbig Ae stars,
HD\,144432 and HD\,144668, for which
magnetic fields are measured at the $\sim$1.6\,$\sigma$
and $\sim$2.5\,$\sigma$ level respectively.
Further, we report the presence of circular polarization signatures in the \ion{Ca}{ii}~K 
line in the V Stokes spectra of HD\,139614 and HD\,144432, which appear unresolved at the 
low spectral resolution achievable with FORS\,1. We suggest that models involving accretion 
of matter from the disk to the star along a global stellar magnetic field of a specific
geometry can account for the observed Zeeman signatures.

           }

\maketitle

\section{Introduction}\label{sec:intro}

It is generally accepted that star formation occurs via accretion.
A number of Herbig Ae stars and classical T\,Tauri stars are surrounded 
by active accretion disks and, probably, most of the excess emission seen at
various wavelength regions can be attributed to the interaction of the disk
with a magnetically active star {\change(e.g.\ Muzerolle et al.\ \cite{Mu04})}. 
This interaction is generally referred to as
magnetospheric accretion.
T\,Tauri stars are solar type
pre-main sequence stars, and from detailed
magnetohydrodynamic models it is expected that their strong magnetic fields funnel material 
from
the disk onto the star and launch a collimated bipolar outflow {\change (e.g. Shu et al.\ \cite{Sh00})}. 

Recent observations of the disk properties of intermediate mass Herbig Ae stars 
suggest a close parallel to T\,Tauri stars, revealing the same size range of
the disks, similar
optical surface brightness and similar structure consisting of inner
dark disk and a bright ring. A large number of T\,Tauri stars are known to
drive micro-jets as well as parsec-scale Herbig-Haro outflows
(McGroarty \& Ray \cite{McGR04}).
Similar outflows have been newly discovered also from intermediate
mass Herbig Ae stars (McGroarty et~al.\ \cite{McG04}).
Spectacular observations in the last years have
been carried out by the Space Telescope Imaging Spectrograph on board of the
Hubble Space Telescope,
providing the first detections in Ly$\alpha$ of micro-jets associated with 
two of the nearest Herbig Ae stars, HD\,163296 and HD\,104237
(Devine et~al.\ \cite{De00}; Grady et~al.\ \cite{Gr04}).
 
Although magnetic fields are believed to play a crucial role in controlling
accretion onto, and winds from, Herbig Ae stars, contrary to the advance
achieved in magnetic studies of T\,Tauri stars, there is still no 
observational evidence demonstrating the strength, extent, and geometry
of their magnetic fields. 
There were several attempts in the past to measure
magnetic fields in these stars (e.g., Glagolevskij \& Chountonov \cite{GC01};
Catala et~al.\ \cite{Ca93}),
but the only detection has been reported for the optically
brightest Herbig Ae star HD\,104237 viewed with the disk close to face-on
(Donati et~al.\ \cite{Do97}).
Using the Anglo-Australian Telescope with the 
UCL Echelle Spectrograph and a visitor polarimeter a marginal
circular polarization signature has been observed in metallic lines 
indicating a longitudinal magnetic field of $\sim$50\,G.

There are certain reasons 
why the detection of magnetic fields in Herbig Ae stars is so difficult.
The young intermediate mass stars as well as more massive stars tend to be surrounded by 
large 
amounts of circumstellar gas and dust, making it hard to measure magnetic fields at 
visual wavelengths. 
{\change A number of spectral lines show circumstellar absorptions with complex
profiles, and some lines appear in emission or exhibit P\,Cyg-type profiles.
The optical spectra of Herbig Ae stars contain
fewer absorption lines compared to cooler stars, and the rotational
velocities of A stars as a class tend to be higher. 
In addition, Herbig Ae stars are less common than T\,Tauri stars, since the
IMF favors the production of low mass stars, and, therefore, the sample of Herbig Ae stars 
includes
rather faint objects.} 
Yet, the lack of magnetic field
detections demonstrates that magnetic fields in Herbig Ae stars should be rather weak, of
the order of a few hundred Gauss and less, 
and a definite detection of such weak 
magnetic fields in faint objects certainly requires the collecting areas of
large aperture telescopes of 8-10\,m class coupled with highly efficient
spectrographs to achieve
a signal-to-noise ratio of at least 1,000.

\section{Observations}\label{sec:obs}

We have used FORS\,1 (FOcal Reducer low dispersion Spectrograph) at the VLT
on September 18 and 21 2003
in spectropolarimetric mode to measure magnetic fields in
three Herbig Ae stars, HD\,139614, HD\,144432 and HD\,144668. 
The multi-mode instrument FORS\,1 is equipped with
polarization analyzing optics comprising super-achromatic half-wave and
quarter-wave phase retarder plates and a Wollaston prism with a beam
divergence of 22$^{\prime\prime}$ in standard resolution mode. We used the GRISM\,600B to cover all H 
Balmer lines from H$_\beta$ to the Balmer jump, and the narrowest available slit width 
of 0$\farcs$4 to obtain a spectral resolving power of R$\sim$2000. 
The assessment of the longitudinal magnetic field using the FORS\,1 spectra
is achieved by measuring the
circular polarization of opposite sign induced in the wings of broad lines,
such as Balmer lines, by the Zeeman effect.
{\change The measurement of circular polarization in magnetically sensitive lines is the most direct mean
of detecting magnetic fields on stellar surfaces.
The errors of the measurement of the polarization in all Balmer lines
from H$_\beta$ to H$_{16}$ have been determined from photon counting 
statistics and have been converted to errors of field measurements. For each Herbig Ae star we 
took eight sub-exposures  
with an integration time of about one minute each (half a minute for the brightest
star HD\,144668) with the retarder waveplate oriented at 
different angles. The spectropolarimetric 
capability of the FORS\,1 
instrument in combination with the large light collecting power of the VLT allowed us to achieve 
a S/N ratio of up to 1400 per pixel around 4600\,\AA{} in the one--dimensional spectrum,
Usually, once in a few months an additional star with a 
well-defined strong longitudinal field is observed to check that the
instrument is functioning properly.}

Details of the observing technique and of the measurement of magnetic fields in the 
FORS\,1 spectra can be found elsewhere (e.g., Bagnulo et~al.\ \cite{Ba02};
Hubrig et~al.\ \cite{Hu04a}).
The major advantage of using low-resolution spectropolarimetry with FORS\,1
is that polarization can be detected also in relatively fast rotators as we
measure the field in the hydrogen Balmer lines and not in metallic lines.

\begin{table*}
\caption{
\label{tab:overview}
Basic data of the studied Herbig Ae stars.}
\begin{center}
\begin{tabular}{rlccccccccr}
\hline\noalign{\smallskip}
\multicolumn{1}{c}{HD} &
\multicolumn{1}{c}{Other} &
\multicolumn{1}{c}{$V$} &
\multicolumn{1}{c}{Sp.\ Type} &
\multicolumn{1}{c}{\teff} &
\multicolumn{1}{c}{\logg} &
\multicolumn{1}{c}{\vsini}&
\multicolumn{1}{c}{Ref.}&
\multicolumn{1}{c}{$(V-L)_{\rm obs}$} &
\multicolumn{1}{c}{$P$} &
\multicolumn{1}{c}{$\left<{\cal B}_z\right>$} \\
\hline\noalign{\smallskip}
139614 & CD-27 10778& 8.2&A7Ve& 8250& 4.5&13&1&$\sim$2${^m}.0$&$\sim$0.1--0.5\% & $-$$450\pm93$\,G\\
144432 & CD-42 10650&8.2&A9Ve & 7750& 4.5&54&1&$\sim$2${^m}.0$&$\sim$0.1--0.5\% & $-$$94\pm60$\,G\\
144668 & HR\,5999&7.0&A7IVe& 7800& 3.5-4.0&180&2&$3{^m}.0$--$3{^m}.5$&$\sim$0.5--1.3\% & $-$$118\pm48$\,G\\
\hline\noalign{\smallskip}
\end{tabular}
\end{center}

(1) Meeus et~al.\ (\cite{Me98}); 
(2) Grady et~al.\ (\cite{Gr94})\\
\end{table*}

\begin{figure*}
\centering
\begin{tabular}{c}
\includegraphics[width=0.33\textwidth]{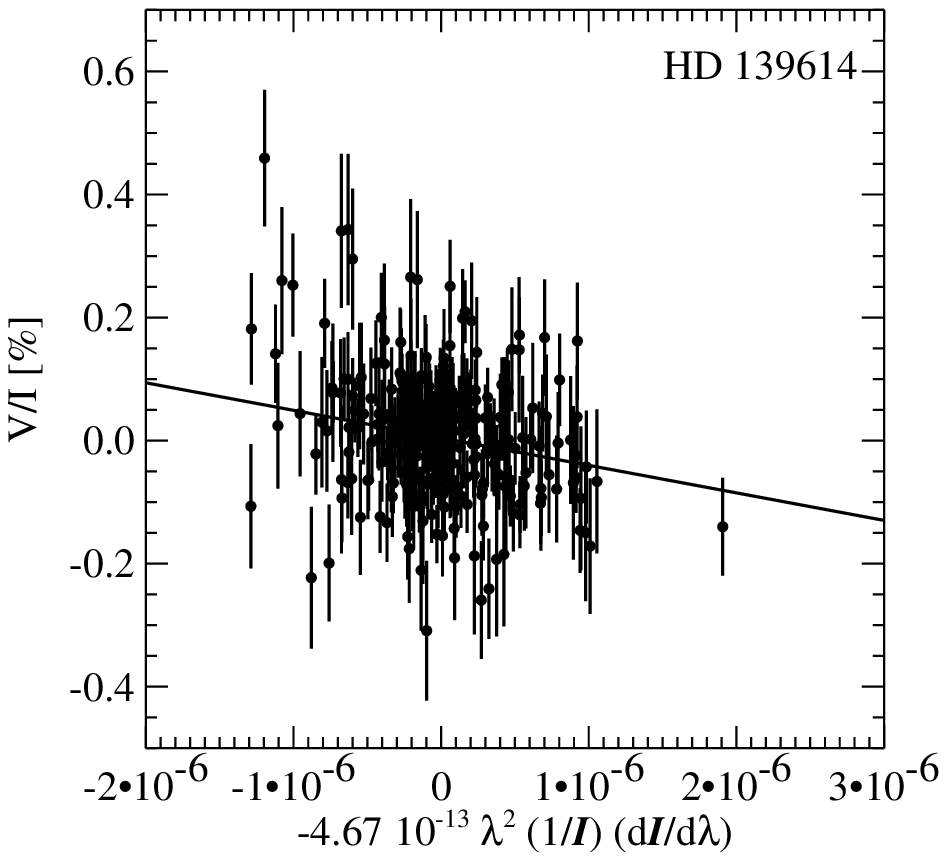}
\includegraphics[width=0.33\textwidth]{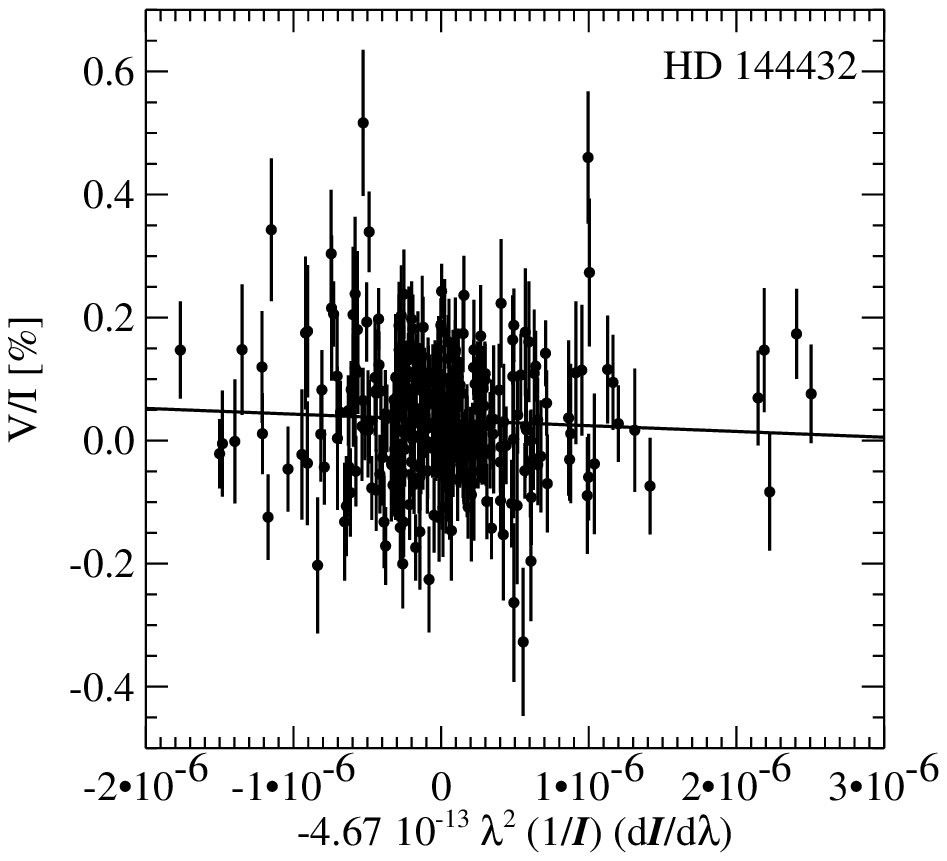}
\includegraphics[width=0.33\textwidth]{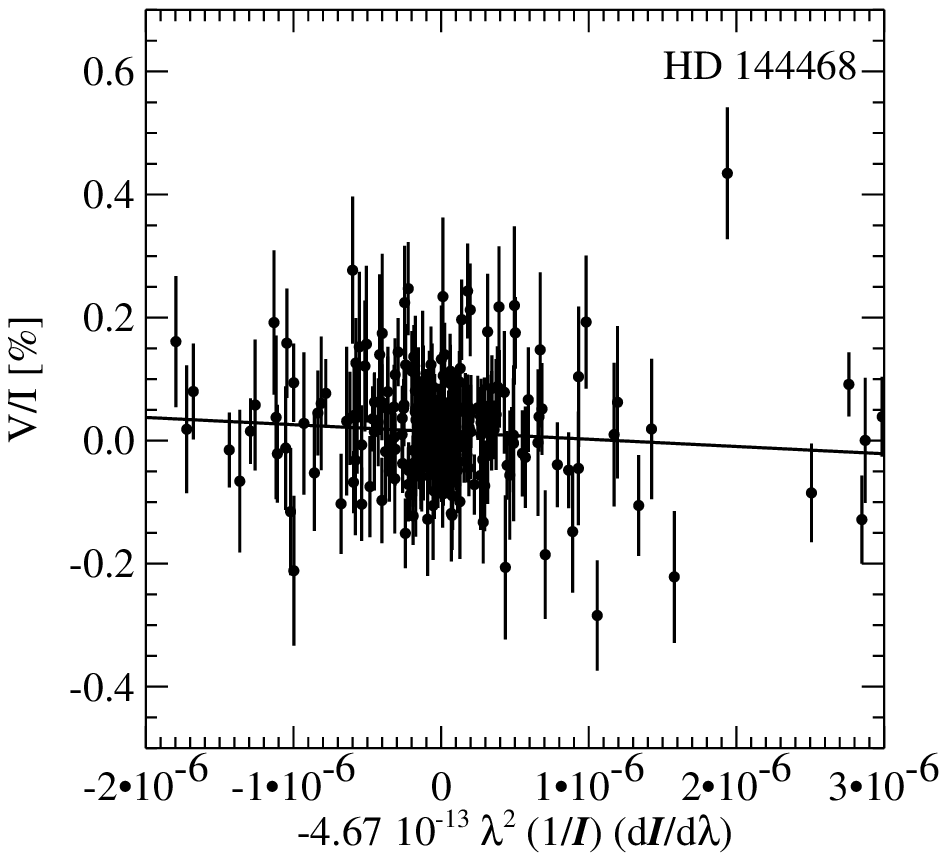}
\end{tabular}
\caption{Regression detection of a $-450\pm93$\,G magnetic field in HD\,139614
and non-detections in HD\,144432 and HD\,144668.
}
\label{fig:Fits}
\end{figure*}

\section{Results and discussion}\label{sec:disc}
The basic data of the studied Herbig Ae stars are presented in Table~\ref{tab:overview}.
The first eight columns indicate, 
in order, the HD number of the star, another identifier, the visual magnitude, the spectral 
type, the stellar parameters and their sources.
The other columns list the infrared colour excess $(V-L)_{obs}$, the linear 
polarization $P$ (Yudin \cite{Yu00}) and our determination of the
mean longitudinal magnetic field $\left<{\cal B}_z\right>$.
{\change The mean longitudinal magnetic field is the average over the stellar hemisphere visible at 
the time of observation of the component of the magnetic field parallel to the line of sight, 
weighted by the local emergent spectral line intensity.
It is diagnosed from the slope of a linear regression of $V/I$
versus the quantity
$-g_{\rm eff} \Delta\lambda_z \lambda^2 \frac{1}{I} \frac{{\mathrm d}I}{{\mathrm d}\lambda} \left<B_z\right> + V_0/I_0$
(Fig.\,\ref{fig:Fits}).
This procedure is described in 
detail
by Bagnulo et al.\ (\cite{Ba02}) and Hubrig et al.\ (\cite{Hu04a}).
Our experience
with a study of a large sample of magnetic and non-magnetic Ap and Bp stars revealed  that this
regression technique is very robust and that detections with $B_z > 3\,\sigma_z$ result only
for stars possessing magnetic fields (Hubrig et al.\ \cite{Hu04b}).}

The differences in the measured infrared colour excess and linear polarization between 
these stars may be interpreted in terms of different inclination of their rotation axis to 
the line of sight, and thus 
of their circumstellar (CS) disks with respect to the observer. 
HD\,139614 and HD\,144432 are very likely observed close to the rotational
pole (face-on for the CS disk; Meeus et~al.\ \cite{Me98})
{\change whereas HD\,144668 is possibly observed at an intermediate angle of $68^\circ>i>45^\circ$
(Natta \& Whitney \cite{NW00})}.
The pre-main sequence nature of HD\,139614 and HD\,144432 
has been studied several years ago (Dunkin et~al.\ \cite{Du97}).
Both stars show emission-line characteristics of H$\alpha$, Na~I D and
He~I 5876\,\AA{} indicating accretion activity and winds.
No age determination has been found in the literature
for HD\,139614.
Although this star has been mentioned in several studies as a Vega-type star
(Sylvester et~al.\ \cite{Sy96}),
it is certainly a member of the Herbig Ae group,
considering its spectral energy distribution in the infrared
and its emission line characteristics.
For HD\,144432, 
the age between 1 and 3 million years has been determined quite recently
(Perez et~al.\ \cite{Pe04}).
The star HD\,144668 is much younger with an age of about 0.5 million years
(van den Ancker et~al.\ \cite{vdA98}).
A search for a magnetic field was previously only performed for HD\,144432.
Using 26 metallic lines Glagolevskij \& Chountonov (\cite{gc01}) measured  
$\left<{\cal B}_z\right>$=$-$$800\pm600$\,G. 
This measurement was severely  hampered by the rather large width of the
measured spectral lines as this star is rotating relatively fast.

\begin{figure}
\centering
{\includegraphics[width=0.4\textwidth]{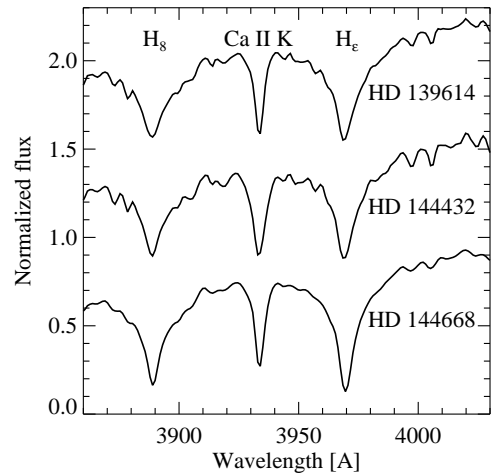}}
\caption{
Normalized Stokes I spectra of the three Herbig Ae stars.
The individual spectra are displaced by 0.65 with respect to each other.
}
\label{fig:StokesI2}
\end{figure}

Spectra of the studied Herbig Ae stars in
integral light in the spectral region around the \ion{Ca}{ii}~K
line and close-by H Balmer lines
are presented in Fig.\,\ref{fig:StokesI2}. 
From our data,
we derive for the star HD\,139614, viewed with the disk close to face-on,
the mean longitudinal
field $\left<{\cal B}_z\right>$=$-$$450\pm93$~G. This is the largest
mean longitudinal magnetic field ever diagnosed for a Herbig Ae star.
For the stars HD\,144432
and HD\,144668 we measure respectively
$\left<{\cal B}_z\right>$=$-$$94\pm60$\,G and
$\left<{\cal B}_z\right>$=$-$$118\pm48$\,G.

\begin{figure}
\centering
{\includegraphics[width=0.4\textwidth]{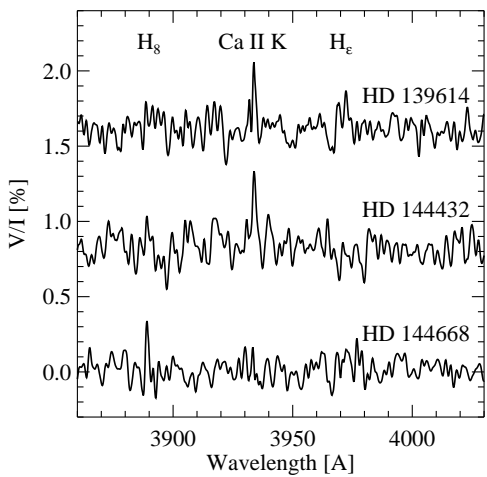}}
\caption{
Stokes V/I spectra of the studied stars around the \ion{Ca}{ii} doublet.
The individual spectra are shifted by 0.8 with respect to each other.
}
\label{fig:StokesV2}
\end{figure}


In the Stokes V spectra of HD\,139614 and HD\,144432 the interesting but
rather unexpected fact is the presence
of circular polarization signatures in the \ion{Ca}{ii}~K line at 
$\lambda\,3933.7$\,\AA{}, which appear unresolved at the low spectral
resolution achievable with FORS\,1 (Fig.\,\ref{fig:StokesV2}).
The line \ion{Ca}{ii}~H at $\lambda\, 3968.5$\,\AA{} is blended with 
the Balmer line H$\epsilon$.
As already noted, Herbig Ae stars are surrounded by circumstellar disks. 
{\change Recent magnetospheric accretion models for these stars assume a dipolar magnetic 
field
geometry and accreting gas from a circumstellar disk falling ballistically 
along the field lines to the stellar surface (Muzerolle et al. 2004).
According to a previous study of the \ion{Ca}{ii} emissivity from dense envelopes
near Herbig stars which used photoionization calculations (Hamann \& Persson \cite{HP92}),
it is very likely that
the \ion{Ca}{ii} lines form very near the photosphere.
Probably, models involving accretion of circumstellar matter from the disk to the
star along a global stellar magnetic field of a specific
geometry can account for the Zeeman signatures observed in the \ion{Ca}{ii} lines.
We note, however, that the actual geometry of magnetic fields in Herbig Ae stars is 
completely unknown.}
Obviously, further spectropolarimetric observations at higher resolution in the 
spectral region around the \ion{Ca}{ii}~H and K lines as well as hydrogen Balmer lines
would provide an essential clue for the understanding of the accretion
mechanism and the geometry of magnetic fields in these stars. 

The observations of the Herbig Ae stars were carried out within a 
framework of the study of evolution of magnetic fields in stars across
the upper main sequence. It has been frequently mentioned in the literature that
they are potential progenitors of the magnetic Ap stars
(e.g., Stepien \& Landstreet \cite{SL02}; Catala \cite{Ca03}).
A detection of magnetic fields in Herbig Ae stars is especially important in
view of our recent results that Ap magnetic stars of mass
below 3\,$M_\odot$ are significantly evolved
and concentrated towards the centre of the main-sequence band, and
practically no magnetic star of mass
below 3\,$M_\odot$ can be found close to the zero-age main sequence
(Hubrig et~al.\ \cite{Hu00}; Hubrig et~al.\ \cite{Hu04b}).
The  search for magnetic fields and the study of their structure in the
pre-main sequence
counterparts to the magnetic Ap stars is a crucial step towards understanding the origin of the
magnetic fields in stars of intermediate mass.

To properly assess the role of magnetic 
fields in the star formation process it is 
important to carry out magnetic field studies of a large sample of 
Herbig Ae stars and to try to establish the magnetic field strength and the 
magnetic field geometry by monitoring over several rotation cycles. 
In T\,Tauri stars, time series of circular polarization measurements 
of spectral lines formed throughout the photosphere and those in accretion regions 
show that the
photosphere as a whole lacks globally organized magnetic fields, but accretion 
regions have highly ordered fields of the order of a few kG (Johns-Krull et~al.\ \cite{JK03}). 
Such contrary behaviour is usually reconciled
by a notion that accreting material is loaded onto magnetic field lines at 
several stellar radii, where the dipolar component dominates higher order 
components of a more complicated surface geometry.
The net longitudinal magnetic field in T\,Tauri stars with an upper limit of 200\,G
is modulated on rotational time scales, implying significant obliquity 
between the magnetic and rotational axes
(Johns-Krull et~al.\ \cite{JK03}).
The current magnetohydrodynamic models of Herbig Ae stars should be 
modified by inclusion of magnetic
fields of proper strength and geometry and we should look forward towards future 
magnetohydrodynamic calculations
of processes taking place in the accretion regions and the photosphere of these 
pre-main sequence stars.

\end{document}